\documentclass[aps,nofootinbib,preprint,superscriptaddress]{revtex4}%
\usepackage{hyperref}
\usepackage{amsmath}
\usepackage{amsfonts}
\usepackage{amssymb}
\usepackage{graphicx}
\usepackage{color}%
\providecommand{\U}[1]{\protect\rule{.1in}{.1in}}
\providecommand{\U}[1]{\protect\rule{.1in}{.1in}}

\begin{document}
\title{The exact $SL(K+3,\mathbb{C})$ symmetry of string theory}
\author{Sheng-Hong Lai}
\email{xgcj944137@gmail.com}
\affiliation{Department of Electrophysics, National Yang Ming Chiao Tung University,
Hsinchu, Taiwan, R.O.C.}
\author{Jen-Chi Lee}
\email{jcclee@cc.nctu.edu.tw}
\affiliation{Department of Electrophysics, National Yang Ming Chiao Tung University,
Hsinchu, Taiwan, R.O.C.}
\affiliation{Center for Theoretical and Computational Physics (CTCP), National Yang Ming
Chiao Tung University, Hsinchu, Taiwan, R.O.C.}
\author{Yi Yang}
\email{yiyang@mail.nctu.edu.tw}
\affiliation{Department of Electrophysics, National Yang Ming Chiao Tung University,
Hsinchu, Taiwan, R.O.C.}
\affiliation{Center for Theoretical and Computational Physics (CTCP), National Yang Ming
Chiao Tung University, Hsinchu, Taiwan, R.O.C.}

\begin{abstract}
By using on-shell recursion relation of string scattering amplitudes (SSA), we
show that \textit{all} $n$-point SSA of the open bosonic string theory can be
expressed in terms of the Lauricella functions. This result extends the
previous exact $SL(K+3,%
\mathbb{C}
)$ symmetry of the $4$-point Lauricella SSA (LSSA) of three tachyons and one
arbitrary string states to the whole tree-level open bosonic string theory.
Moreover, we present three applications of the $SL(K+3,%
\mathbb{C}
)$ symmetry on the SSA. They are the solvability of all $n$-point SSA in terms
of four-tachyon amplitudes, the existence of iteration relations among
residues of a given SSA so as to soften its hard scattering behavior and
finally the re-derivation of infinite linear relations among hard SSA
\cite{CHLTY2}.

\end{abstract}
\maketitle

\section{Introduction}

Symmetry has long been considered as an important property of physical laws
for both classical and quantum physics. Historically, symmetry was used to be
thought of as one of the direct consequences of equation of motion (EOM) of a
physical law before the developments of general relativity and Yang-Mills
theory. However, it was soon realized that, in a quantum field theory (QFT)
for example, symmetry principle was even more fundamental than the EOM itself.
Indeed, it was symmetry which determined the form of the interaction or EOM of
a physical law (symmetry dictates interaction). More importantly, for the case
of electrodynamics, in contrast to classical physics, \textit{only} in quantum
theory \cite{yang} can the QED $U(1)$ gauge symmetry be identified and used to
derive Ward identities which secure QED as a consistent renormalizable QFT.

In QFT, for example QCD, one usually considers interactions with up\ to
four-point couplings whose forms are fixed by the symmetry principle. In
addition, symmetry principle can be used to derive Slavnov-Taylor identities
which relate different couplings. This is in contrast to the four-fermion
model in the weak interaction which is not a consistent renormalizable QFT. In
string theory, on the contrary, one is given a set of rules through quantum
consistency of the extended string which was used to fix the forms of
interactions or vertices to calculate perturbative on-shell string scattering
amplitudes (SSA). Moreover, instead of up to four-point couplings in QCD, one
encounters $n$-point couplings with arbitrary $n$ which correspond to the
infinite number of degrees of freedom in the spectrum of string theory. One
crucial issue of string theory is thus to identify symmetry of the theory and
uses it to relate these infinite number of couplings of particles with
arbitrary higher masses and spins.

To identify the exact symmetry group (or even a smaller subgroup) of string
theory is much more complicated than that of a QFT. This is because in string
theory one needs to deal with infinite number of massive couplings or vertices
instead of up to four in a typical QFT. The well-known $E_{8}^{2}$ and
$SO(32)$ symmetries of the $10D$ Heterotic string are symmetries of Yang-Mills
couplings in the massless sector only. In a series of recent papers, the
present authors calculated a subset of exact $4$-point SSA, namely, amplitudes
of three tachyons and one arbitrary string states, and expressed them in terms
of the $D$-type Lauricella functions \cite{LLY2}. In addition, it was shown
that these Lauricella SSA (LSSA) can be expressed in terms of the basis
functions in the \textit{infinite} dimensional representation of the
$SL(K+3,C)$ group \cite{Group}. It is important to note that instead of finite
dimensional representation of a compact Lie group, here we encounter an
infinite dimensional representation of a noncompact Lie group. For any fixed
positive integer $K$, we have infinite number of LSSA in the $SL(K+3,C)$
representations \cite{slkc}. Moreover, it was further shown that there existed
$K+2$ recurrence relations among the $D$-type Lauricella functions. These
recurrence relations can be used to reproduce the Cartan subalgebra and simple
root system of the $SL(K+3,C)$ group with rank $K+2$. As a result, the
$SL(K+3,C)$ group with its corresponding stringy Ward identities (recurrence
relations) can be used to solve \cite{solve} all the LSSA and express them in
terms of one amplitude. See the recent review paper \cite{LSSA}.

As an important application of this solvability in the hard string scattering
limit, the $SL(K+3,C)$ symmetry group of the LSSA can be used to reproduce
\cite{LLY2} infinite linear relations with constant coefficients among all
hard SSA and solve the ratios among them. These high energy behaviors of
string theory \cite{GM,GM1} were first conjectured by Gross \cite{Gross} and
later corrected and proved \cite{ChanLee,ChanLee2,CHLTY2,CHLTY1} by the method
of decoupling of zero norm states (ZNS) \cite{Lee,lee-Ov,LeePRL}. See the
review papers \cite{review,over}. Since the decoupling of ZNS and thus the
infinite linear relations in the hard scattering regime persist to all string
loop orders, we conjecture that the $SL(K+3,C)$ symmetry at string-tree level
proposed in this letter is also valid for string loop amplitudes. One early
attempt using the so-called bracket relations to identify stringy symmetries
can be found in \cite{Moore1}. However, neither Lie algebra structure nor the
complete recurrence relations to solve all SSA in terms of one amplitude were
identified. Nevertheless, it is still an interesting problem to find the
connections between the bracket relations and the $K+2$ Lauricella recurrence
relations associated with the $SL(K+3,C)$ group.

In this letter, we will apply the string theory extension
\cite{bcfw5,bcfw4,bcfw3,stringbcfw} of field theory BCFW on-shell recursion
relations \cite{bcfw1,bcfw2} to show that the $SL(K+3,C)$ symmetry group of
the $4$-point LSSA persists for general $n$-point SSA with arbitrary higher
point couplings in string theory. We thus have shown that, at least at string
tree level, the $SL(K+3,C)$ symmetry is an exact symmetry of the whole bosonic
string theory.

One main effort of this letter is to show that all residues of SSA in the
string theory on-shell recursion prescription can be expressed in terms of the
four-point LSSA. We thus conclude that all $n$-point SSA of the bosonic string
theory form an infinite dimensional representation of the $SL(K+3,C)$
symmetry. Indeed, we can use mathematical induction, together with the
on-shell recursion and the shifting principle, to show that all $n$-point SSA
can be expressed in terms of the LSSA.

On the other hand, since the LSSA of three tachyons and one arbitrary string
states can be rederived \cite{LLYT} from the deformed cubic string field
theory (SFT) \cite{Taejin}, it is conjectured that the proposed $SL(K+3,C)$
symmetry in this letter can be hidden in Witten SFT.

\section{The 4-point SSA}

We begin with a brief review of the LSSA of three tachyons and one arbitrary
string states in the $26D$ open bosonic string theory and its associated
$SL(K+3,C)$ group. The general states at mass level $M^{2}=2(N-1)$, where
$N=\sum_{X}\sum_{n>0}nr_{n}^{X}\geq0$ is an integer representing the mass
level, are of the following form \cite{LSSA}
\begin{equation}
\left\vert r^{X}\right\rangle =\prod_{X}\left(  \prod_{n>0}\left(  \alpha
_{-n}^{X}\right)  ^{r_{n}^{X}}\right)  |0,k\rangle\label{state}%
\end{equation}
where $X$ labels the momentum, longitudinal and transverse polarizations on
the $\left(  2+1\right)  $-dimensional scattering plane.

The 4-point LSSA associated to the above string state Eq.(\ref{state}) can be
calculated to be \cite{LSSA}%
\begin{equation}
A_{4}=B\left(  -\frac{t}{2}-1,-\frac{s}{2}-1\right)  F_{D}^{(K)}\left(
-\frac{t}{2}-1;R_{n}^{X};\frac{u}{2}+2-N;\tilde{Z}_{n}^{X}\right)  \prod
_{X}\left(  \prod_{n=1}\left[  -(n-1)!k_{3}^{X}\right]  ^{r_{n}^{X}}\right)
\label{st}%
\end{equation}
where $B(a,b)$ is the Beta function with $\left(  s,t\right)  $ being the
usual Mandelstam variables, $k_{i}^{X}$ is the momentum of the $i$th string
state projected on the $X$ polarization, and%
\begin{equation}
K=\sum_{X}\underset{\{\text{for all }r_{j}^{X}\neq0\}}{\sum j}%
\end{equation}
is an integer depending on the polarization.

The $D$-type Lauricella function $F_{D}^{(K)}$ in Eq.(\ref{st}) is one of the
four extensions of the Gauss hypergeometric function to $K$ variables and is
defined to be%
\begin{equation}
F_{D}^{(K)}\left(  \alpha;\beta_{1},...,\beta_{K};\gamma;x_{1},...,x_{K}%
\right)  =\sum_{n_{1},\cdots,n_{K}=0}^{\infty}\frac{\left(  \alpha\right)
_{n_{1}+\cdots+n_{K}}}{\left(  \gamma\right)  _{n_{1}+\cdots+n_{K}}}%
\frac{\left(  \beta_{1}\right)  _{n_{1}}\cdots\left(  \beta_{K}\right)
_{n_{K}}}{n_{1}!\cdots n_{K}!}x_{1}^{n_{1}}\cdots x_{K}^{n_{K}}%
\end{equation}
where $(\alpha)_{n}=\alpha\cdot\left(  \alpha+1\right)  \cdots\left(
\alpha+n-1\right)  $ is the Pochhammer symbol.

For the multi-tensor cases, there are new terms with finite number of
contractions among $\partial^{n}X...$ and $\partial^{m}X...$, and one obtains
more $D$-type Lauricella functions with different values of $K$. In general, a
state at mass level $N$ with $M^{2}=2\left(  N-1\right)  ,$ and $N=\sum
_{r>0}rN_{r},$ is of the form (it is understood, for example, that the state
$\epsilon_{1}^{(1)}\cdot\alpha_{-1}\epsilon_{1}^{(2)}\cdot\alpha_{-1}$ means
$\epsilon_{\mu\nu}\alpha_{-1}^{\mu}\alpha_{-1}^{\nu}$) $\left\vert
P\right\rangle =\prod_{r>0}\prod_{\sigma=1}^{N_{r}}\frac{\varepsilon
_{r}^{\left(  \sigma\right)  }\cdot\alpha_{-r}}{\sqrt{N_{r}!r^{N_{r}}}%
}|0,k\rangle$ where $\varepsilon_{r}^{\left(  \sigma\right)  }$ are
polarizations with $\sigma=1,\cdots N_{r}$ for each operator $\alpha_{-r} $.
The 4-point SSA with $i=1,2,3,4$ can be calculated to be $(z_{ij}=z_{i}%
-z_{j})$%
\begin{align}
A_{4}  &  =\int_{0}^{1}dz_{2}z_{2}^{k_{1}\cdot k_{2}}\left(  1-z_{2}\right)
^{k_{2}\cdot k_{3}}\nonumber\\
&  \cdot\sum_{\left\{  \varepsilon_{r_{i}}^{\left(  \sigma_{i}\right)
}\right\}  }\left[  \prod\limits_{i=1}^{4}\prod\limits_{\left\{  r_{i}%
,\sigma_{i}\right\}  }\left(  \sum_{j\neq i}\frac{\varepsilon_{r_{i}}^{\left(
\sigma_{i}\right)  }\cdot k_{j}}{z_{ji}^{r_{i}}}\right)  \cdot\prod
\limits_{i<j=2}^{4}\prod\limits_{\left\{  r_{i},\sigma_{i};r_{j},\sigma
_{j}\right\}  }\frac{\varepsilon_{r_{i}}^{\left(  \sigma_{i}\right)
}\varepsilon_{r_{j}}^{\left(  \sigma_{j}\right)  }}{z_{ji}^{r_{i}}%
z_{ij}^{r_{j}}}\right]  _{z_{1}=0,z_{3}=1,z_{4}\rightarrow\infty} \label{line}%
\end{align}
where\ the configurations $\left\{  \varepsilon_{r_{i}}^{\left(  \sigma
_{i}\right)  }\right\}  $\ satisfy%
\begin{equation}
\prod\limits_{i=1}^{4}\prod\limits_{\left\{  r_{i},\sigma_{i}\right\}
}\varepsilon_{r_{i}}^{\left(  \sigma_{i}\right)  }\cdot\prod\limits_{i<j=2}%
^{4}\prod\limits_{\left\{  r_{i},\sigma_{i};r_{j},\sigma_{j}\right\}  }\left(
\varepsilon_{r_{i}}^{\left(  \sigma_{i}\right)  }\varepsilon_{r_{j}}^{\left(
\sigma_{j}\right)  }\right)  =\prod_{i=1}^{4}\prod_{r_{i}>0}\prod_{\sigma
_{i}=1}^{N_{r_{i}}}\varepsilon_{r_{i}}^{\left(  \sigma_{i}\right)  },
\end{equation}
which ensures the multi-linear condition. For each configuration $\left\{
\varepsilon_{r_{i}}^{\left(  \sigma_{i}\right)  }\right\}  $, it is
straightforward to transform Eq.(\ref{line}) to the \textit{sum} of standard
integral form of the Lauricella functions.

\section{The $SL\left(  K+3,C\right)  $ symmetry}

To obtain the $SL(K+3,C)$ symmetry of the LSSA, it is important to note that
one can define the basis functions \cite{slkc}%
\begin{align}
&  f_{ac}^{b_{1}\cdots b_{K}}\left(  \alpha;\beta_{1},\cdots,\beta_{K}%
;\gamma;x_{1},\cdots,x_{K}\right) \nonumber\\
&  =B\left(  \gamma-\alpha,\alpha\right)  F_{D}^{\left(  K\right)  }\left(
\alpha;\beta_{1},\cdots,\beta_{K};\gamma;x_{1},\cdots,x_{K}\right)  a^{\alpha
}b_{1}^{\beta_{1}}\cdots b_{K}^{\beta_{K}}c^{\gamma}, \label{id2}%
\end{align}
so that the LSSA in Eq.(\ref{st}) can be written as \cite{Group}
\begin{equation}
A_{4}=f_{11}^{-(n-1)!k_{3}^{X}}\left(  -\frac{t}{2}-1;R_{n}^{X},;\frac{u}%
{2}+2-N;\tilde{Z}_{n}^{X}\right)  .
\end{equation}
We then introduce the $(K+3)^{2}-1$ generators $\mathcal{E}_{ij}$ of
$SL(K+3,C)$ group \cite{slkc,Group}%
\begin{equation}
\left[  \mathcal{E}_{ij},\mathcal{E}_{kl}\right]  =\delta_{jk}\mathcal{E}%
_{il}-\delta_{li}\mathcal{E}_{kj};\text{ \ }1\leqslant i,j\leqslant K+3.
\end{equation}
These are $1$ $E^{\alpha}$, $K$ $E^{\beta_{k}}(k=1,2\cdots K)$, $1$
$E^{\gamma}$,$1$ $E^{\alpha\gamma}$, $K$ $E^{\beta_{k}\gamma}$ and $K$
$E^{\alpha\beta_{k}\gamma}$ which sum up to $3K+3$ raising generators. There
are also $3K+3$ lowering operators. In addition, there are $K\left(
K-1\right)  $ $E_{\beta_{p}}^{\beta_{k}}$ and $K+2$ $\ J$ , $\left\{
J_{\alpha},J_{\beta_{k}},J_{\gamma}\right\}  $, the Cartan subalgebra. In sum,
the total number of generators are $2(3K+3)+K(K-1)+K+2=(K+3)^{2}-1$
\cite{Group}.

For the general $4$-point LSSA, it is straightforward to calculate the
operation of these generators on the basis functions ($k=1,2,...K$)
$f_{ac}^{b_{1}\cdots b_{K}}\left(  \alpha;\beta_{1},\cdots,\beta_{K}%
;\gamma;x_{1},\cdots,x_{K}\right)  $, and show the $SL(K+3,C)$ symmetry
\cite{Group}. For the cases of higher point $(n\geq5)$ LSSA, one encounters
sum of products of the Lauricella functions (\textit{extended LSSA} or simply
LSSA)%
\begin{equation}
\text{Residue of }n\text{-point }LSSA\sim\sum\text{coefficient}\prod
(\text{single tensor }4\text{-point }LSSA) \label{pro}%
\end{equation}
where the residue will be defined in Eq.(\ref{res}). Therefore, one needs to
deal with product representations of $SL(K+3,C)$.

\section{Residues of the n-poing KN amplitudes}

\begin{figure}[ptb]
\centering
\includegraphics[
height=1.0153in,
width=7.1788in
]{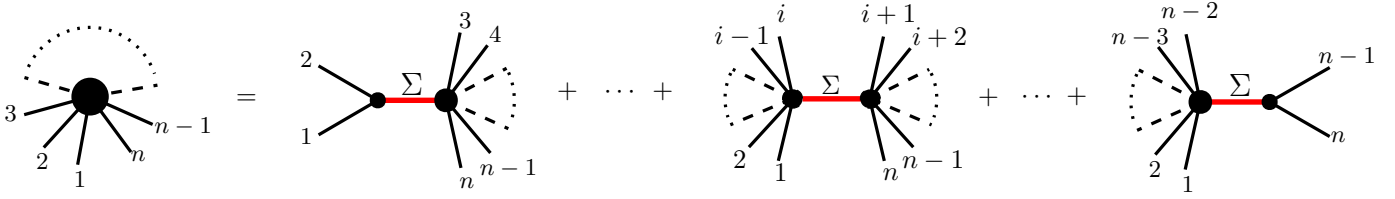}
\caption{Residues of KN amplitudes}
\label{recursion}%
\end{figure}

After showing that all $4$-point SSA can be expressed as the Lauricella
functions, we consider the general $n$-point ($n\geq5$) SSA now. The key is to
apply the string theory extension of field theory BCFW on-shell recursion
relations. We first consider the KN amplitude. Applying the BCFW deformation
$\hat{k}_{1}\left(  z\right)  =k_{1}+zq$ and $\hat{k}_{n}\left(  z\right)
=k_{n}-zq$ with $q^{2}=k_{1}\cdot q=k_{n}\cdot q=0$, the locations of the
poles $z_{m,M}$ are given by solutions of $\left(  \hat{k}_{1}+k_{2}%
+....+k_{m}\right)  ^{2}+2(M-1)=0,$ $M=0,1,2...$. The $n$-point Koba-Nielsen
(KN) amplitude can then be written as \cite{stringbcfw}%
\begin{equation}
A_{n}^{KN}=\underset{m=2}{\overset{n-2}{\sum}}\underset{M=0}{\overset{\infty
}{\sum}}\frac{2R_{m+1,n-m+1}^{M}}{\left(  k_{1}+k_{2}+...+k_{m}\right)
^{2}+2(M-1)}, \label{res}%
\end{equation}
which is represented in Fig.\ref{recursion}.

\begin{figure}[ptb]
\centering
\includegraphics[
height=2.3212in,
width=4.6337in
]{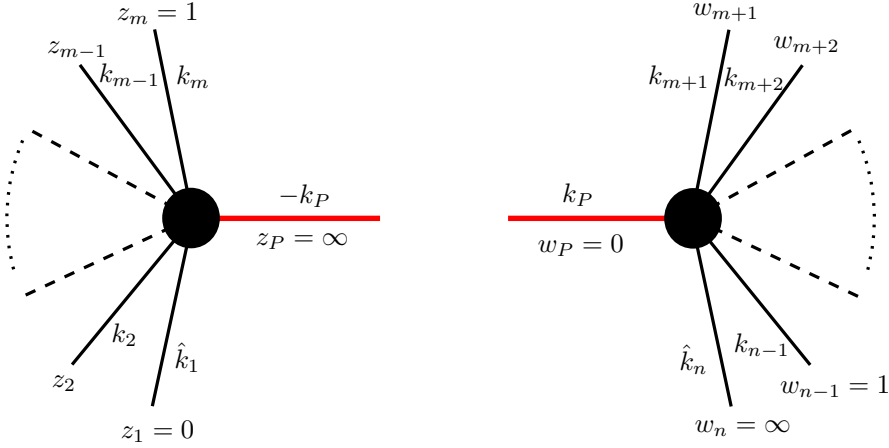}
\caption{The left and right subamplitudes}%
\label{subamp}%
\end{figure}

It turns out that the residue $R_{m+1,n-m+1}^{M}$ in Eq.(\ref{res}) can be
calculated and expressed in terms of the subamplitudes \cite{LLY3}%
\begin{equation}
R_{m+1,n-m+1}^{M}=\sum_{P}A_{m+1}^{L}\left(  1,\cdots,m,-P\right)
A_{n-m+1}^{R}\left(  P,m+1,\cdots n\right)  \label{RAA}%
\end{equation}
where%
\begin{align}
&  A_{m+1}^{L}\left(  1,\cdots,m,-P\right) \nonumber\\
&  =\int_{0}^{1}dz_{m-1}\cdots\int_{0}^{z_{3}}dz_{2}\underset{1\leq j<i\leq
m}{\prod}\left(  z_{i}-z_{j}\right)  ^{k_{i}\cdot k_{j}}\prod_{r=1}%
\frac{\left[  \underset{\sigma=1}{\overset{N_{r}}{%
{\displaystyle\prod}
}}\epsilon_{r}^{(\sigma)}\cdot\underset{2\leq i\leq m}{\sum}k_{i}\left(
z_{i}\right)  ^{r}\right]  }{\sqrt{N_{r}!r^{N_{r}}}},\label{LL}\\
&  A_{n-m+1}^{R}\left(  P,m+1,\cdots n\right) \nonumber\\
&  =\int_{0}^{1}dw_{n-2}\cdots\int_{0}^{w_{m+2}}dw_{+1}\underset{P=m\leq
j<i\leq n-1}{\prod}\left(  w_{i}-w_{j}\right)  ^{k_{i}\cdot k_{j}}\prod
_{r=1}\frac{\left[  \underset{\sigma=1}{\overset{N_{r}}{%
{\displaystyle\prod}
}}(-\epsilon_{r}^{(\sigma)})\cdot\underset{m+1\leq i\leq n-1}{\sum}%
k_{i}\left(  \frac{1}{w_{i}}\right)  ^{r}\right]  }{\sqrt{N_{r}!r^{N_{r}}}}
\label{RR}%
\end{align}
with $k_{P}=\hat{k}_{1}+\underset{i=2}{\overset{m}{\sum}}k_{i}$. It is
understood that $k_{1}$ in Eq.(\ref{LL}) should be replaced by $\hat{k}_{1}$.
The results in Eq.(\ref{LL}), Eq.(\ref{RR}) are consistent with a direct
calculation from KN amplitude \cite{LLY3}. The diagram representation of
Eq.(\ref{RAA}) is given in Fig.\ref{subamp}.

In extending the field theory BCFW to the string theory BCFW above, one
encountered difficulty of poles at infinity. To resolve the problem, the
authors of \cite{bcfw4} used the pomeron vertex operators in the Regge regime
constructed by BPST \cite{RR6} to show the vanishing of the string amplitudes
at large complex momenta provided that the Regge behavior of the deformed
amplitude is power law falloff $\mathcal{M}(z)\sim z^{n+1+\frac{t}{2}}$ (for
sufficiently negative $t$) where $n$ depends on the mass levels. Indeed, the
\textit{universal} power law falloff behavior $V\sim\lbrack ik_{2}%
\cdot\partial X]^{1+\frac{t}{2}}$ \cite{Tan} for arbitrary massive pomeron
vertex and their associated universal power law falloff Regge amplitudes
$A_{Regge}$ $\sim s^{1+\frac{t}{2}}$ \cite{KLY} had all been shown to be
independent of the mass levels of the vertex ($n=0$).

\section{Expressing n-point SSA in terms of LSSA}

In the last subsection, we have shown in Eq.(\ref{res}) and Eq.(\ref{RAA}%
)\ that the $n$-point KN amplitude can be expressed in terms of product of
lower-point sub-amplitudes. The next step is to consider $n$-point SSA with
tensor legs. To do it, we introduce the following shifting principle:

\textit{Shifting principle} : If the $n$-point KN amplitude $A_{n}^{KN}$ can
be expressed in terms of the LSSA, then one can use the shifting method to
calculate all $n$-point SSA $A_{n}$ with tensor legs (excited string states)
and express them in terms of the LSSA ($\left(  z_{1},z_{n-1},z_{n}\right)
=\left(  0,1,\infty\right)  $)%

\begin{equation}
A_{n}(k_{i}\cdot k_{j};\zeta_{k})\sim%
{\displaystyle\sum}
c(k_{i};\zeta_{k})A_{n}^{KN}(k_{i}\cdot k_{j}\rightarrow k_{i}\cdot
k_{j}-a_{ij}). \label{kkll}%
\end{equation}
In Eq.(\ref{kkll}), $k_{i}\cdot k_{j}$ with $1\leq i<j\leq n-1$ $(k_{i}\cdot
k_{j}\neq k_{1}\cdot k_{n-1})$ are kinematic variables of the $n$-point SSA
which sums up to $\frac{(n-1)(n-2)}{2}-1=$ $\frac{n(n-3)}{2}$, $a_{ij}$ are
shifting integers and $\zeta_{k}$ are polarizations of the excited string states.

For example, for a $5$-point SSA with four tachyons and one vector ($\left(
z_{1},z_{4},z_{5}\right)  =\left(  0,1,\infty\right)  $) \cite{LLY3}%
\begin{align}
&  A_{5}\left(  \zeta_{1},k_{1};k_{2};k_{3};k_{4};k_{5}\right) \nonumber\\
&  =\int_{0}^{1}dz_{3}\int_{0}^{z_{3}}dz_{2}z_{2}^{k_{2}\cdot k_{1}}%
z_{3}^{k_{3}\cdot k_{1}}\left(  1-z_{3}\right)  ^{k_{4}\cdot k_{3}}\left(
1-z_{2}\right)  ^{k_{4}\cdot k_{2}}\left(  z_{3}-z_{2}\right)  ^{k_{3}\cdot
k_{2}}(-\zeta_{1})\cdot\left[  \frac{k_{2}}{z_{2}}+\frac{k_{3}}{z_{3}}%
+\frac{k_{4}}{1}\right] \nonumber\\
&  =A_{5}^{(1)}+A_{5}^{(2)}+A_{5}^{(3)}. \label{three}%
\end{align}
\ where one reads $\left(  a_{12}=1,a_{ij\neq12}=0\right)  $ for $A_{5}^{(1)}%
$, $\left(  a_{13}=1,a_{ij\neq13}=0\right)  $ for $A_{5}^{(2)}$ and $a_{ij}=0$
for $A_{5}^{(3)}$. The result can be shown to be a sum of LSSA.
Mathematically, the calculation of all three terms of Eq.(\ref{three}) are
similar to that of the $5$-point KN amplitude with shifting some appropriate
kinematic variables. Although the calculation of $5$-point SSA with higher
tensor legs is very lengthy, it is trivial by the shifting method adopted in
the calculation of Eq.(\ref{three}) to see that all of them are LSSA.

We see that while the string on-shell recursion relation can be used to reduce
higher point KN amplitudes to the lower SSA and express them in terms of the
LSSA, the shifting method can be used to express SSA with tensor legs in terms
of the LSSA.

In general, we can use mathematical induction, together with the on-shell
recursion and the shifting principle, to show that all $n$-point SSA can be
expressed in terms of the LSSA. The procedure goes as following. We assume
that all $k$-point SSA ($k\leq n-1$) are LSSA, and we want to prove that all
$n$-point SSA are LSSA. To prove this, we first apply the on-shell recursion
to express the residue of $n$-point KN amplitude calculated in Eq.(\ref{RAA})
in terms of the lower point ($k\leq n-1$) SSA which were assumed to be LSSA.
So the $n$-point KN amplitude is a LSSA. We can then apply the shifting
principle to show that all $n$-point SSA including the $n$-point KN amplitude
are LSSA. This completes the proof.

\begin{figure}[ptb]
\centering
\includegraphics[
height=1.2592in,
width=6.1056in
]{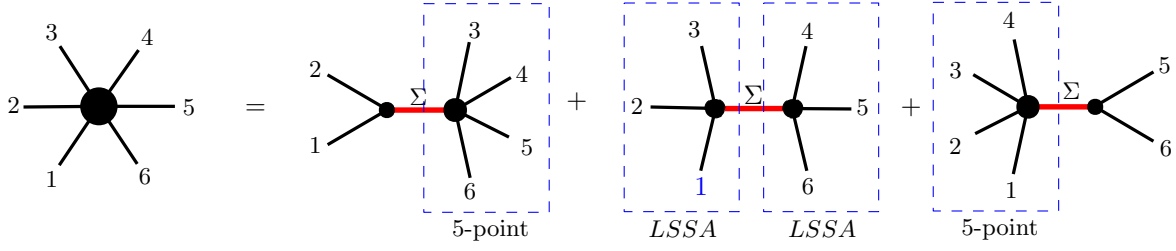}\caption{Expressing
the 6-point KN amplitude in terms of the LSSA by the first recursion}%
\label{6point}%
\end{figure}

\begin{figure}[ptb]
\centering
\includegraphics[
height=1.286in,
width=4.4797in
]{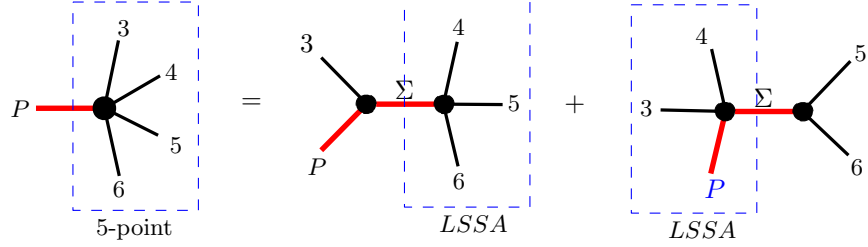}\caption{Expressing
the 6-point KN amplitude in terms of the LSSA by the second recursion}%
\label{6point2}%
\end{figure}

Finally, we use the example of $6$-point KN amplitude to demonstrate its LSSA
form. See Fig.\ref{6point} and Fig.\ref{6point2}. We note that to show the
$6$-point KN amplitude is a LSSA, one needs only do $1$-step recursion to
express it in terms of the lower $5$-point and $4$-point amplitudes as was
shown in Fig.\ref{6point}. Since we have shown that all $5$-point and
$4$-point SSA are LSSA, the $6$-point KN amplitude is a LSSA. However, to
explicitly calculate the LSSA form of the $6$-point KN amplitude, one needs to
do the second recursion on the first and the third diagrams of
Fig.\ref{6point} as was shown in Fig.\ref{6point2}, and the calculation will
be very lengthy as there are two higher excited string states (two heavy
lines) involved. In general, the residues of the $n$-point KN amplitude can be
expressed as a LSSA with application of up to $(recursion)^{n-4}$.

$\boldsymbol{Applications.}$

\section{Applications}

After showing that all open bosonic SSA can be expressed in terms of the LSSA,
we will demonstrate three applications of the associated $SL(K+3,C)$ Symmetry
of LSSA on SSA:

\subsection{Solvability}

The first one is to use $SL(K+3,C)$ group to solve all SSA and express them in
terms of one amplitude. We first show that there exist $K+2$ recurrence
relations for the $D$-type Lauricella functions. To do this, one begins with
the type-$1$ Appell functions. For the case of $K=2$, the $D$-type Lauricella
functions $F_{D}^{(K)}\left(  \alpha;\beta_{1},...,\beta_{K};\gamma
;x_{1},...,x_{K}\right)  $ reduce to the type-$1$ Appell functions
$F_{1}\left(  \alpha;\beta_{1},\beta_{2};\gamma,x,y\right)  $, and one has $4$
known recurrence relations. It was then shown that one can generalize the
$4=2+2$ fundamental recurrence relations of the Appell functions $F_{1}$ and
prove the following $K+2$ recurrence relations for the $D$-type Lauricella
functions $(m=1,2,...,K)$ \cite{Group}%
\begin{align}
\left(  \alpha-\underset{i}{\sum}\beta_{i}\right)  F_{D}^{(K)}-\alpha
F_{D}^{(K)}\left(  \alpha+1\right)  +\beta_{1}F_{D}^{(K)}\left(  \beta
_{1}+1\right)  +...+\beta_{K}F_{D}^{(K)}\left(  \beta_{K}+1\right)   &  =0,\\
\gamma F_{D}^{(K)}-\left(  \gamma-\alpha\right)  F_{D}^{(K)}\left(
\gamma+1\right)  -\alpha F_{D}^{(K)}\left(  \alpha+1;\gamma+1\right)   &
=0,\\
\gamma F_{D}^{(K)}+\gamma(x_{m}-1)F_{D}^{(K)}\left(  \beta_{m}+1\right)
+(\alpha-\gamma)x_{m}F_{D}^{(K)}\left(  \beta_{m}+1,;\gamma+1\right)   &  =0
\end{align}
where for simplicity we have omitted those arguments of $F_{D}^{(K)}$ which
remain the same in the relations. Moreover, these recurrence relations can be
used to reproduce the Cartan subalgebra and simple root system of the
$SL(K+3,C)$ group with rank $K+2$ \cite{Group}. With the Cartan subalgebra and
the simple roots, one can easily write down the whole Lie algebra of the
$SL(K+3,C)$ group. So one can construct the $SL(K+3,C)$ Lie algebra from the
recurrence relations and vice versa%
\begin{equation}
SL(K+3,C)\text{ Lie algebra }\Longleftrightarrow\text{ Recurrence relations of
Lauricella.}%
\end{equation}

\bigskip The next step is to use the above recurrence relations to deduce the
following key recurrence relation \cite{solve}%
\begin{equation}
x_{j}F_{D}^{(K)}\left(  \beta_{i}-1\right)  -x_{i}F_{D}^{(K)}\left(  \beta
_{j}-1\right)  +\left(  x_{i}-x_{j}\right)  F_{D}^{(K)}=0,
\end{equation}
which can be repeatedly used to decrease the value of $K$ and reduce all the
Lauricella functions $F_{D}^{(K)}$ in the LSSA to the Gauss hypergeometric
functions $F_{D}^{(1)}=$ $_{2}F_{1}(\alpha,\beta,\gamma,x)$. One can further
reduce the Gauss hypergeometric functions by deriving a multiplication theorem
for them, and then solve \cite{solve} all the LSSA in terms of one single
amplitude, say the four tachyon amplitude. See the recent review \cite{LSSA}.
For the cases of higher point functions in Eq.(\ref{pro}), all amplitudes can
be solved and expressed in terms of sum of products of the four tachyon
amplitudes. One of the reason of this solvability is that all $\beta_{J}$ in
the Lauricella functions of the LSSA take very special values, namely,
nonpositive integers.

\subsection{Iteration Relations}

The second application is to use the $SL(K+3,C)$ symmetry to show the
existence of iteration relations among residues of a given SSA so as to soften
its hard scattering behavior. It is well known that one can express the
Veneziano amplitude in terms of a series of simple pole terms with residues%
\begin{equation}
A_{4}=\overset{\infty}{\underset{n=0}{\sum}}-\frac{(\alpha(t)+1)(\alpha
(t)+2)\cdots(\alpha(t)+n)}{n!}\frac{1}{\alpha(s)-n}%
\end{equation}
where $\alpha(t)=\alpha^{\prime}t+\alpha(0)$ with $\alpha^{\prime}=1/2$ and
$\alpha(0)=1$ \cite{GSW}. Instead of naive $\sum\frac{1}{n}$ divergence,
$A_{4}$ above behaves as exponential fall-off in the hard scattering limit due
to the iteration relations of the residues among each pole term. We will find
generalization of the iteration relations among residues in $A_{4}$ for higher
point KN amplitudes in the following. We note that because of the solvability
discussed in subsection $\boldsymbol{(A)}$ or the $SL(K+3,C)$ symmetry of the
LSSA, we expect relations among various residues $M=0,1,2\ldots$of a given
$R_{m+1,n-m+1}^{M}$ which is a sum of LSSA for a fixed $M$ and $(m,n)$.
Indeed, if we define
\begin{align}
F_{M}\left(  x_{1},\cdots,x_{n}\right)   &  \equiv\underset{%
\begin{array}
[c]{c}%
a_{1}+\cdots+a_{n}=M\\
a_{1},\cdots,a_{n}=0
\end{array}
}{\sum}\frac{\left(  T_{1}\right)  _{a_{1}}}{a_{1}!}\cdots\frac{\left(
T_{n}\right)  _{a_{n}}}{a_{n}!}x_{1}^{a_{1}}\cdots x_{n}^{a_{n}}\nonumber\\
&  =\underset{\left\{  N_{r}\right\}  \text{,with fixed }M=\underset{r}{\sum
}rN_{r}}{\sum}\prod_{r=1}\left(  \frac{1}{N_{r}!r^{N_{r}}}\left[  T_{1}%
x_{1}^{r}+\cdots+T_{n}x_{n}^{r}\right]  ^{N_{r}}\right)  , \label{id1}%
\end{align}
one can prove the second equality in Eq.(\ref{id1}). We can now express the
residue $R_{m+1,n-m+1}^{M}$ in terms of $F_{M}$%

\begin{align}
&  R_{m+1,n-m+1}^{M}\left(  \hat{1},2,\cdots,m,m+1,\cdots,n-1,\hat{n}\right)
\nonumber\\
&  =\int_{0}^{1}dz_{m-1}\cdots\int_{0}^{z_{3}}dz_{2}\int_{0}^{1}dw_{n-2}%
\cdots\int_{0}^{w_{m+2}}dw_{m+1}\underset{1\leq j<i\leq m}{\prod}\left(
z_{i}-z_{j}\right)  ^{k_{i}\cdot k_{j}}\nonumber\\
&  \times\underset{P=m\leq j<i\leq n-1}{\prod}\left(  w_{i}-w_{j}\right)
^{k_{i}\cdot k_{j}}F_{M}\left(  \frac{z_{j}}{w_{i}}\right)  \label{kkk1}%
\end{align}
where $j=2,\cdots,m;i=m+1,\cdots,n-1$ and $z_{m}=w_{n-1}=1$, and show the
following iteration relation \newline%
\begin{equation}
F_{M}=\frac{1}{M}\underset{N=0}{\overset{M-1}{\sum}}F_{N}\left[
\underset{2\leq j\leq m,m+1\leq i\leq n-1}{\sum}\left(  -k_{j}\cdot
k_{i}\right)  \left(  \frac{z_{j}}{w_{i}}\right)  ^{M-N}\right]  , \label{kkk}%
\end{equation}
which expresses $F_{M}$ in terms of $F_{M-1}$, $F_{M-2}$, $....$, $F_{1}$,
$F_{0}\equiv1$. For illustration, we give one explicit example here. The $n=4$
term in the Veneziano amplitude $A_{4}$ can be written as
\begin{equation}
A_{4}=\cdots+\left(  \underset{\alpha_{-4}}{\underbrace{\frac{T}{4}}%
}+\underset{\alpha_{-3}\alpha_{-1}}{\underbrace{\frac{T^{2}}{1!1!}\frac
{1}{1^{1}}\frac{1}{3^{1}}}}+\underset{\alpha_{-2}^{2}}{\underbrace{\frac
{T^{2}}{2!}\frac{1}{2^{2}}}}+\underset{\alpha_{-1}^{2}\alpha_{-2}%
}{\underbrace{\frac{T^{3}}{2!1!}\frac{1}{1^{2}2^{1}}}}+\underset{\alpha
_{-1}^{4}}{\underbrace{\frac{T^{4}}{4!}\frac{1}{1^{4}}}}\right)  \int_{0}%
^{1}dzz^{k_{1}\cdot k_{2}+4}+\cdots\label{mass}%
\end{equation}
where $T=-k_{2}\cdot k_{3}$. On the other hand, the explicit form of the
residue $R_{4,3}^{4}$ of the $5$-point KN amplitude can be calculated to be%

\begin{align}
R_{4,3}^{4}  &  =\frac{1}{4}\int_{0}^{1}dz_{2}z_{2}^{\hat{k}_{1}\cdot k_{2}%
}\left(  1-z_{2}\right)  ^{k_{2}\cdot k_{3}}\left[  -k_{3}\cdot k_{4}%
-(-k_{2}\cdot k_{4})z_{2}^{4}\right] \nonumber\\
&  +\frac{1}{1!1!}\left(  \frac{1}{3}\right)  \left(  \frac{1}{1}\right)
\int_{0}^{1}dz_{2}z_{2}^{\hat{k}_{1}\cdot k_{2}}\left(  1-z_{2}\right)
^{k_{2}\cdot k_{3}}\left[  -k_{3}\cdot k_{4}-(-k_{2}\cdot k_{4})z_{2}%
^{3}\right]  \left[  -k_{3}\cdot k_{4}-(-k_{2}\cdot k_{4})z_{2}\right]
\nonumber\\
&  +\frac{1}{2!}\left(  \frac{1}{2}\right)  ^{2}\int_{0}^{1}dz_{2}z_{2}%
^{\hat{k}_{1}\cdot k_{2}}\left(  1-z_{2}\right)  ^{k_{2}\cdot k_{3}}\left[
-k_{3}\cdot k_{4}-(-k_{2}\cdot k_{4})z_{2}^{2}\right]  ^{2}\nonumber\\
&  +\frac{1}{2!1!}\left(  \frac{1}{1}\right)  ^{2}\left(  \frac{1}{2}\right)
\int_{0}^{1}dz_{2}z_{2}^{\hat{k}_{1}\cdot k_{2}}\left(  1-z_{2}\right)
^{k_{2}\cdot k_{3}}\left[  -k_{3}\cdot k_{4}-(-k_{2}\cdot k_{4})z_{2}\right]
^{2}\left[  -k_{3}\cdot k_{4}-(-k_{2}\cdot k_{4})z_{2}^{2}\right] \nonumber\\
&  +\frac{1}{4!}\left(  \frac{1}{1}\right)  ^{4}\int_{0}^{1}dz_{2}z_{2}%
^{\hat{k}_{1}\cdot k_{2}}\left(  1-z_{2}\right)  ^{k_{2}\cdot k_{3}}\left[
-k_{3}\cdot k_{4}-(-k_{2}\cdot k_{4})z_{2}\right]  ^{4}. \label{R5}%
\end{align}
We see that the coefficients of each of the $5$ terms in Eq.(\ref{R5}) and
Eq.(\ref{mass}) are the same! Furthermore, we have the correspondence
$\alpha_{-l}\Longrightarrow\left[  -k_{3}\cdot k_{4}-(-k_{2}\cdot k_{4}%
)z_{2}^{l}\right]  $. Similarly, $R_{4,3}^{0}$, $R_{4,3}^{1}$, $R_{4,3}^{2}$
and $R_{4,3}^{3}$ (and the corresponding $R_{3,4}^{M}$) can all be expressed
in terms of the LSSA. Moreover, because of the solvability or the $SL(K+3,C)$
symmetry of the LSSA, we expect relations among various residues $R_{4,3}^{0}%
$, $R_{4,3}^{1}$, $R_{4,3}^{2}$, $R_{4,3}^{3}$, $R_{4,3}^{4}\cdots$ and
$R_{3,4}^{0}$, $R_{3,4}^{1}$, $R_{3,4}^{2}$, $R_{3,4}^{3}$, $R_{3,4}^{4}%
\cdots$ of the $5$-point KN amplitude as was given in Eq.(\ref{kkk}) and
Eq.(\ref{kkk1}).

\subsection{Linear Relations}

It was first observed that for each fixed mass level $N$ with $M^{2}=2(N-1)$,
the following states are of leading order in energy at the hard scattering
limit \cite{CHLTY2,CHLTY1}
\begin{equation}
\left\vert N,2m,q\right\rangle \equiv(\alpha_{-1}^{T})^{N-2m-2q}(\alpha
_{-1}^{L})^{2m}(\alpha_{-2}^{L})^{q}|0,k\rangle. \label{Nmq}%
\end{equation}
One important application of the LSSA presented in Eq.(\ref{st}) is to
reproduce \cite{LLY2} infinite linear relations among all hard $4$-point SSA
and solve the ratios among them \cite{CHLTY2,CHLTY1}
\begin{equation}
\frac{A^{(N,2m,q)}}{A^{(N,0,0)}}=\left(  -\frac{1}{M}\right)  ^{2m+q}\left(
\frac{1}{2}\right)  ^{m+q}(2m-1)!!. \label{ratio}%
\end{equation}
These high energy behaviors of string theory \cite{GM,GM1} were first
conjectured by Gross \cite{Gross} and later corrected and proved
\cite{ChanLee,ChanLee2,CHLTY2,CHLTY1} by the method of decoupling of zero norm
states (ZNS) \cite{Lee,lee-Ov,LeePRL}.

Since the linear relations obtained by the decoupling of ZNS are valid order
by order and share the same forms for all orders in string perturbation
theory, one expects that there exists \textit{stringy symmetry} of the theory
associated with the ratios in Eq.(\ref{ratio}). In fact, there is a simple
analogy from the ratios of the nucleon-nucleon scattering processes in
particle physics $(a)$ \ $p+p\rightarrow d+\pi^{+},(b)$ \ $p+n\rightarrow
d+\pi^{0}$ and $(c)$ \ $n+n\rightarrow d+\pi$, which can be calculated to be
(ignore the mass difference between proton and neutron) $A_{a}:A_{b}%
:A_{c}=1:\frac{1}{\sqrt{2}}:1$ from $SU(2)$ isospin symmetry. Two such
symmetry groups were suggested recently to be the $SL(5,C)$ group in the Regge
string scattering limit \cite{review,over} and the $SL(4,C)$ group in the
Non-relativistic string scattering limit \cite{review,over}. Moreover, it was
shown that the ratios in Eq.(\ref{ratio}) can be extracted from the Regge SSA
\cite{review,over}. With the discovery of the LSSA, we now understand that the
ratios in Eq.(\ref{ratio}) are associated with the exact $SL(K+3,C)$ group
\cite{LLY2}. Finally, For the cases of higher point $(n\geq5)$ functions in
Eq.(\ref{pro}), it is conjectured that there exist hard scattering regimes for
which the linear relations persist and the ratios can be solved accordingly.

\section{Conclusion}

xWe conclude the discussion in this letter with the following analogy of
fundamental symmetries between field theory and string theory%
\begin{align}
SU(2)\text{ symmetry }  &  \Longrightarrow\text{ Yang-Mills \textit{field}
theory,}\\
\text{Bosonic open \textit{string} theory}  &  \Longrightarrow\text{
}SL(K+3,C)\text{ symmetry,}\\
SU(2)\text{ slavnov-Taylor identities }  &  \Longleftrightarrow
SL(K+3,C)\text{ recurrence relations.}%
\end{align}

This work is supported in part by the Ministry of Science and Technology
(MoST) and S.T. Yau center of National Yang Ming Chiao Tung University (NYCU), Taiwan.


\begin{thebibliography}{99}                                                                                               %


\bibitem {yang}C.N.Yang, "Hermann Weyl's contribution to physics" in
\textit{Hermann Weyl}, ed. K. Chandlagrasekharan, Springer-Verlag 1980.

\bibitem {LLY2}Sheng-Hong Lai, Jen-Chi Lee, and Yi~Yang. \newblock The
Lauricella functions and exact string scattering amplitudes. \newblock {\em
Journal of High Energy Physics}, 2016(11):62, 2016.

\bibitem {Group}S. H. Lai, J. C. Lee and Yi Yang, "The $SL(K+3,C)$ Symmetry of
the Bosonic String Scattering Amplitudes", Nucl. Phys. B 941 (2019) 53-71.

\bibitem {slkc}Willard Miller~Jr. \newblock Symmetry and separation of
variables. Addison-Wesley, Reading, Massachusetts, \newblock1977.

\bibitem {solve}S. H. Lai, J. C. Lee and Yi Yang, "Solving Lauricella String
Scattering Amplitudes through recurrence relations", JHEP 09 (2017) 130.

\bibitem {LSSA}S. H. Lai, J. C. Lee and Yi Yang, "Recent developments of the
Lauricella string scattering amplitudes and their exact $SL(K+3,C)$ Symmetry",
Symmetry 13 (2021) 454, arXiv:2012.14726 [hep-th].

\bibitem {GM}David~J Gross and Paul~F Mende. \newblock {The high-energy
behavior of string scattering amplitudes}. \newblock {\em Phys. Lett. B},
197(1):129--134, 1987.

\bibitem {GM1}David~J Gross and Paul~F Mende. \newblock {String theory
beyond the Planck scale}. \newblock {\em Nucl. Phys. B}, 303(3):407--454, 1988.

\bibitem {Gross}David~J. Gross. \newblock {High-Energy Symmetries of String
Theory}. \newblock {\em Phys. Rev. Lett.}, 60:1229, 1988.

\bibitem {ChanLee}Chuan-Tsung Chan and Jen-Chi Lee. \newblock {Stringy
symmetries and their high-energy limits}. \newblock {\em Phys. Lett. B},
611(1):193--198, 2005.

\bibitem {ChanLee2}Chuan-Tsung Chan and Jen-Chi Lee. \newblock {Zero-norm
states and high-energy symmetries of string theory}. \newblock {\em Nucl.
Phys. B}, 690(1):3--20, 2004.

\bibitem {CHLTY2}Chuan-Tsung Chan, Pei-Ming Ho, Jen-Chi Lee, Shunsuke
Teraguchi, and Yi~Yang. \newblock {High-energy zero-norm states and
symmetries of string theory}. \newblock {\em Phys. Rev. Lett.}, 96(17):171601, 2006.

\bibitem {CHLTY1}Chuan-Tsung Chan, Pei-Ming Ho, Jen-Chi Lee, Shunsuke
Teraguchi, and Yi~Yang.
\newblock {Solving all 4-point correlation functions for bosonic open string
	theory in the high-energy limit}. \newblock {\em Nucl. Phys. B},
725(1):352--382, 2005.

\bibitem {Lee}Jen-Chi Lee. \newblock New symmetries of higher spin states in
string theory. \newblock {\em Physics Letters B}, 241(3):336--342, 1990.

\bibitem {lee-Ov}Jen-Chi Lee and Burt~A Ovrut.
\newblock {Zero-norm states and enlarged gauge symmetries of the closed bosonic
	string with massive background fields}. \newblock {\em Nucl. Phys. B},
336(2):222--244, 1990.

\bibitem {LeePRL}Jen-Chi Lee. \newblock {Decoupling of degenerate
positive-norm states in string theory}. \newblock {\em Phys. Rev. Lett.},
64(14):1636, 1990.

\bibitem {review}Jen-Chi Lee and Yi~Yang. \newblock Review on high energy
string scattering amplitudes and symmetries of string theory. \newblock {\em
arXiv preprint arXiv:1510.03297}, 2015.

\bibitem {over}Jen-Chi Lee and Yi~Yang. \newblock Overview of high energy
string scattering amplitudes and symmetries of string theory. \newblock {\em
Symmetry}, 11(8):1045, 2019.

\bibitem {Moore1}Gregory Moore. \newblock {Symmetries of the bosonic string
S-matrix}. \newblock {\em arXiv preprint hep-th/9310026}, 1993.

\bibitem {bcfw5}R. Boels, K.J. Larsen, N. A. Obers and Marcel Vonk,
\textquotedblleft MHV, CSW and BCFW: field theory structures in string theory
amplitudes\textquotedblright\ JHEP 0811 (2008) 015 [arXiv:0808.2598 [hep-th]].

\bibitem {bcfw4}C. Cheung, D. O'Connell and B. Wecht, \textquotedblleft BCFW
Recursion Relations and String Theory,\textquotedblright\ JHEP 1009 (2010) 052
[arXiv:1002.4674 [hep-th]].

\bibitem {bcfw3}R. H. Boels, Daniele Marmiroli and N. A. Obers,
\textquotedblleft On-shell Recursion in String Theory,\textquotedblright\ JHEP
1010 (2010) 034 [arXiv:1002.5029 [hep-th]].

\bibitem {stringbcfw}Yung-Yeh Chang, Bo Feng, Chih-Hao Fu, Jen-Chi Lee,Yihong
Wang and Yi Yang, "A note on on-shell recursion relation of string
amplitudes", JHEP 02 (2013) 028.

\bibitem {bcfw1}R. Britto, F. Cachazo and B. Feng, \textquotedblleft New
recursion relations for tree amplitudes of gluons,\textquotedblright\ Nucl.
Phys. B 715 (2005) 499 [hep-th/0412308].

\bibitem {bcfw2}R. Britto, F. Cachazo, B. Feng and E. Witten,
\textquotedblleft Direct proof of tree-level recursion relation in Yang-Mills
theory,\textquotedblright\ Phys. Rev. Lett. 94 (2005) 181602 [hep-th/0501052].

\bibitem {LLYT}Sheng-Hong Lai, Jen-Chi Lee, Taejin Lee and Yi~Yang, Phys.
Lett. B 776 (2018) 150-157.

\bibitem {Taejin}Taejin Lee, Phys. Lett. B 768 (2017) 248.

\bibitem {LLY3}Sheng-Hong Lai, Jen-Chi Lee, and Yi~Yang, "Residues of bosonic
string scattering amplitudes and the Lauricella functions", arXiv:2109.08601 [hep-th].

\bibitem {RR6}Richard~C Brower, Joseph Polchinski, Matthew~J Strassler, and
Chung-I Tan. \newblock {The Pomeron and gauge/string duality}. \newblock
\emph{JHEP}, 2007(12):005, 2007.

\bibitem {Tan}Chih-Hao Fu, Jen-Chi Lee, Chung-I Tan, and Yi~Yang.
\newblock {Recurrence relations of higher spin BPST vertex operators for open
strings}. \newblock {\em Phys. Rev. D}, 88(4):046004, 2013.

\bibitem {KLY}Sheng-Lan Ko, Jen-Chi Lee, and Yi~Yang, "Patterns of high energy
massive string scatterings in the Regge regime", JHEP 06 (2009) 028,
arXiv:0812.4190 [hep-th].

\bibitem {GSW}MB Green, JH Schwarz and E Witten. Superstring theory, v. 1.
Cambridge University,Cambridge, (1987).
\end{thebibliography}

\end{document}